\documentclass[11pt,a4paper]{article}

\title{Deformations of Fermionic Quantum Field Theories and\\ Integrable Models}
\author{Sabina Alazzawi\\\\
\footnotesize{Faculty of Physics, University of Vienna,}\\
\footnotesize{Boltzmanngasse 5, 1090 Wien, Austria}\\
\footnotesize{sabina.alazzawi@univie.ac.at}}
\date{}
\usepackage{geometry}
\usepackage{mathrsfs}
\usepackage{stmaryrd}
\usepackage{hyperref}
\usepackage{amsmath}
\usepackage{amsfonts}
\usepackage{textcomp}
\usepackage{url}
\usepackage[ansinew]{inputenc}
\newtheorem{theorem}{Theorem}[section]
\newtheorem{lemma}[theorem]{Lemma}
\newtheorem{proposition}[theorem]{Proposition}
\newtheorem{corollary}[theorem]{Corollary}
\newtheorem{definition}[theorem]{Definition}
\usepackage{fancyhdr}
\begin{document}

\maketitle
\abstract{Considering the model of a scalar massive Fermion, it is shown that by means of deformation techniques it is possible to obtain all integrable quantum field theoretic models on two-dimensional Minkowski space which have factorizing S-matrices corresponding to two-particle scattering functions $S_2$ satisfying $S_2(0)=-1$. Among these models there is for example the Sinh-Gordon model. Our analysis provides a complement to recent developments regarding deformations of quantum field theories. The deformed model is investigated also in higher dimensions. In particular, locality and covariance properties are analyzed.}
\tableofcontents

\section{Introduction}
Recent developments regarding the construction of quantum field theoretic models have shown that deformation techniques lead to models with non-trivial interaction \cite{BS,BLS,DLM,GG,GG2,GL}. This new approach starts from a well-known model which is subjected to a certain modification. This deformation has to be carried out in such a way that covariance and locality properties of quantum field theory are preserved. The results obtained so far show that this difficult task can be coped with by weakening the locality requirements. The localization regions are in that case so-called \textit{wedges}. A wedge region $W$ in $d$-dimensional Minkowski space is defined as a Poincaré transform of a reference wedge
\begin{equation}\label{W}
W_0:=\{(x_0,\dots,x_{d-1})\in\mathbb{R}^d:x_1>|x_0|\},
\end{equation}
often called the \textit{right wedge}. The notion of wedge-locality therefore refers to the vanishing of commutators of two field operators which are localized in space-like separated wedges.\par
In \cite{GL} G. Lechner presented a deformation method which in two space-time dimensions yields a large class of integrable models. Considering only a single species of particles, the S-matrix of such a model is completely determined by the two-particle scattering function $S_2$. The class of integrable models obtained via deformation techniques in \cite{GL} corresponds to scattering functions with value $+1$ at zero rapidity parameter, i.e. $S_2(0)=+1$. There are, however, very interesting models with $S_2(0)=-1$ which do not fit into the deformation scheme of \cite{GL}. An important example for such a model is the Sinh-Gordon model.\par
In this note, we are mainly concerned with providing a similar analysis that includes integrable models with scattering functions satisfying $S_2(0)=-1$ in the deformation framework and therefore complements the results in \cite{GL}. In particular, our starting point is a quantum field satisfying canonical anticommutation relations instead of canonical commutation relations as in \cite{GL}. The difficulty here consists of the fact that the undeformed model is, except in two spacetime dimensions, non-local from the outset. The re-establishment of the locality property in more than two dimensions has up to now not been accomplished for the deformed model, although in the undeformed case remnants of locality can be found \cite{BS1}.\par
Moreover, we also comment on the bosonic case. In particular, we point out that the analysis in \cite{GL} is unnecessarily restrictive and that a slightly more general result can be obtained (see Lemma \ref{lem2} below).\par
In the next section we introduce the model of a scalar massive Fermion which is then deformed in Section \ref{3}. In addition, the properties of the model obtained in this way are analyzed. Focusing on the two-dimensional case, the deformed model is associated with integrable models. In Section \ref{conclusions} we summarize our findings and point out open questions.

\section{The model of a scalar massive Fermion} \label{fermi}
\subsection{The model}\label{model}
This section is devoted to the specification of the model which describes a scalar massive Fermion. The model at least goes back to the 1960's and can be found in R. Jost's book \cite[p. 103]{J} in connection with weak local commutativity of field operators.\par
In \cite{BS1}, D. Buchholz and S. Summers studied this model in more detail. In particular, they were interested in the degree of nonlocality of the model and investigated if there are any remnants of locality which have physical significance. We recall those findings which are of particular interest for our purposes.\par
To set the stage, let $\mathscr{H}$ denote the antisymmetric Fock space over the one-particle space $\mathscr{H}_{1}$ of a scalar particle of mass $m>0$, that is
\begin{equation}
\mathscr{H}=\oplus_{n=0}^{\infty}\mathscr{H}_{n},\qquad \mathscr{H}_{n}=\mathscr{H}_{1}\wedge\dots\wedge\mathscr{H}_{1},\nonumber
\end{equation}
and $\mathscr{H}_{0}=\mathbb{C}$ consisting of multiples of the vacuum state $\Omega$. Here, we use the following convention
\begin{equation*}
\varphi_1\wedge\dots\wedge\varphi_n:=
\frac{1}{n!}\sum_{\pi}\sigma(\pi)\varphi_{\pi(1)}\otimes\cdots\otimes\varphi_{\pi(n)},\quad\varphi_i\in\mathscr{H}_{1},
\end{equation*}
where the sum is over all permutations $\pi:(1,\dots,n)\mapsto (\pi(1),\dots,\pi(n))$ and $\sigma(\pi)$ is $+1$ if $\pi$ is even and $-1$ if $\pi$ is odd. Furthermore, we use the notation $$\Psi_n(\varphi_1,\dots,\varphi_n)=\varphi_1\wedge\dots\wedge\varphi_n,\quad \varphi_i\in\mathscr{H}_{1}.$$
As usual, we introduce creation and annihilation operators $a^{\#}(\varphi)$ representing the CAR algebra on the Fock space $\mathscr{H}$, i.e. for $\varphi,\psi\in \mathscr{H}_1$ we have
\begin{eqnarray*}
\{a^{*}(\varphi),a^{*}(\psi)\}&=&0\\
\{a(\varphi),a(\psi)\}&=&0\\
\{a(\varphi),a^{*}(\psi)\}&=&\langle \varphi,\psi\rangle\cdot 1.
\end{eqnarray*}
In the following, we shall identify the one-particle space $\mathscr{H}_{1}$ with $L^{2}(\mathbb{R}^{d},d\mu(p))$ where $d\geq 2$ and
\begin{equation*}
d\mu(p):=\omega(\mathbf{p})^{-1}\delta(p^0-\omega(\mathbf{p}))dp,\quad \omega(\mathbf{p})=\sqrt{\mathbf{p}^2+m^2},\quad m>0,\quad p=(p^0,\textbf{p})\in\mathbb{R}^d.
\end{equation*}
In this setting, the Fourier transform $\widetilde{f}$ of a function $f\in\mathscr{S}(\mathbb{R}^d)$,
\begin{equation*}
\widetilde{f}(p):=\int dx f(x) e^{ip\cdot x},\qquad p\cdot x=p^0x^0-\mathbf{p}\mathbf{x},
\end{equation*}
restricted to the positive mass shell ${H_{m}^{+}}=\{p=(p^0,\mathbf{p})\in\mathbb{R}^d: p^0=\omega(\mathbf{p})\}$ is an element of $\mathscr{H}_{1}$, i.e. $\widetilde{f}|_{H_{m}^{+}}\in\mathscr{H}_{1}$. We shall, further, use the notation
\begin{equation*}
f^{\pm}(p):=\widetilde{f}(\pm p)=\int dx f(x) e^{\pm ip\cdot x},\qquad p\in H_{m}^{+}.
\end{equation*}
The scalar product in $\mathscr{H}_{n}$ is given by
\begin{equation*}
\langle\varphi_n|\psi_n\rangle=\int d\mu(p_1)\cdots d\mu(p_n)\overline{\varphi_n(p_1,\dots,p_n)}\psi_n(p_1,\dots,p_n).
\end{equation*}
The action of the annihilation and creation operators is defined by
\begin{eqnarray*}
\left(a(\varphi)\Psi\right)_{n}(p_{1},\dots,p_{n})&:=&\sqrt{n+1}\int d\mu(p) \overline{\varphi(p)}\Psi_{n+1}(p,p_{1},\dots,p_{n}),\\
\left(a^{*}(\varphi)\Psi\right)_{n}(p_{1},\dots,p_{n})&:= &\frac{1}{\sqrt{n}}\sum_{k=1}^{n}(-1)^{k+1}\varphi(p_{k}) \Psi_{n-1}(p_{1},\dots,\hat{p}_{k},\dots,p_{n}),\\
a^{*}(\varphi)\Omega:=\varphi,&\qquad& a(\varphi)\Omega:=0,
\end{eqnarray*}
where $\varphi\in\mathscr{H}_{1}$, $\Psi\in\mathscr{H}$ and $\hat{p}_{k}$ denotes the omission of the variable $p_k$. For further purposes, we introduce the operator-valued distributions $a^{\#}(p)$ such that
\begin{equation*}
a(\varphi)=\int d\mu(p) \overline{\varphi(p)}a(p), \qquad a^{*}(\varphi)=\int d\mu(p) \varphi(p)a^{*}(p).
\end{equation*}
Their action is given by
\begin{subequations}\label{a}
\begin{eqnarray}
\left(a(p)\Psi\right)_{n}(p_{1},\dots,p_{n})&=&\sqrt{n+1}\Psi_{n+1}(p,p_{1},\dots,p_{n}),\\
 \left(a^{*}(p)\Psi\right)_{n}(p_{1},\dots,p_{n})&=&\frac{1}{\sqrt{n}}\sum_{k=1}^{n}(-1)^{k+1}\omega(\mathbf{p}) \delta(\mathbf{p}-\mathbf{p_k})\Psi_{n-1}(p_{1},\dots,\hat{p}_{k},\dots,p_{n}).\nonumber\\
\end{eqnarray}
\end{subequations}
The Poincaré group $\mathcal{P}_+^{\uparrow}$ is represented on $\mathscr{H}$ in the usual manner by the second quantized continuous unitary representation $U$ which leaves $\Omega$ invariant and acts according to
\begin{equation}\label{unitary}
\left(U(a,\Lambda)\Psi\right)_n(p_1,\dots,p_n)=e^{i\sum_{k=1}^{n}p_k\cdot a}\,\Psi_n(\Lambda^{-1}p_1,\dots,\Lambda^{-1}p_n),\qquad (a,\Lambda)\in\mathcal{P}_+^{\uparrow}.
\end{equation}
The joint spectrum of the generators of the translation group $U(a,1)$ is a subset of the forward light cone $\overline{V_+}$, i.e. the translation group satisfies the spectral condition.\par
Proceeding in a standard way, we introduce an operator-valued distribution $\phi:\mathscr{S}(\mathbb{R}^{d})\rightarrow \mathcal{B}(\mathscr{H})$ which is defined by
\begin{equation}\label{feld}
\phi(f):=a^{*}(f^{+})+a(\overline{f^{-}}).
\end{equation}
This field operator obviously satisfies canonical anticommutation relations, in particular,
\begin{equation*}
\{\phi(f),\phi(g)\}=\left(\langle \left(\overline{f}\right)^+|g^+\rangle+\langle \left(\overline{g}\right)^+|f^+\rangle\right)\cdot 1.
\end{equation*}
Furthermore, $\phi$ is a weak solution of the Klein-Gordon equation, $\phi(f)^{*}=\phi(\overline{f})$ and it transforms covariantly under the adjoint action of the unitary representation $U$ of the Poincaré group, i.e.
$$U(a,\Lambda)\phi(f) U(a,\Lambda)^{-1}=\phi(f_{(a,\Lambda)}),\qquad (a,\Lambda)\in\mathcal{P}_+^{\uparrow}, $$
where $f_{(a,\Lambda)}(x):=f\left(\Lambda^{-1}(x-a)\right)$. Note that neither the anticommutator nor the commutator of two field operators $\phi(f)$ and $\phi(g)$ vanishes for spacelike separated supports of $f$ and $g$. This circumstance is consistent with the spin-statistics theorem \cite{J, SW} and expresses the nonlocality of the field $\phi$. In fact, the corresponding net of von Neumann algebras
\begin{equation*}
\mathcal{R}(W):=\{\phi(f):f\in\mathscr{S}(\mathbb{R}^d), \mathrm{supp}f\subset W\}''
\end{equation*}
is maximally nonlocal, i.e.
\begin{equation*}
\mathcal{R}(W)'\cap \mathcal{R}(W')=\mathbb{C}\cdot 1,
\end{equation*}
where $W$ is any wedge region (for a proof see \cite{BS1}). Here  $\mathcal{R}(W)'$ denotes the commutant of $\mathcal{R}(W)$ and $W'$ the spacelike complement of $W$. We shall further denote by $\mathcal{W}$ the set of all wedges. In more than two spacetime dimensions the set $\mathcal{W}$ is given by $\mathcal{W}=\{g\,W_0:g\in\mathcal{P}_+^{\uparrow}\}$ whereas in two dimensions it consists of two disjoint components, namely the translates of $W_0$ on the one hand, see Eqn. (\ref{W}), and the translates of $W_0'=-W_0$ on the other hand.\par
We now introduce an auxiliary field $\widehat{\phi}:\mathscr{S}(\mathbb{R}^{d})\rightarrow \mathcal{B}(\mathscr{H})$
\begin{equation}\label{hat}
\widehat{\phi}(f):=(-1)^{N(N-1)/2}\phi(f)(-1)^{N(N-1)/2}=(a^{*}(f^{+})-a(\overline{f^{-}}))(-1)^{N},
\end{equation}
where $N$ is the particle number operator acting on $\mathscr{H}_{n}$ according to $N|_{\mathscr{H}_n}=n\cdot 1$. The field $\widehat{\phi}$ has the same properties as $\phi$, in particular it is also nonlocal. It turns out, however, that the fields $\phi$ and $\widehat{\phi}$ are relatively local, i.e. the commutator
\begin{equation}\label{relloc}
[\widehat{\phi}(f),\phi(g)]=\left(\langle (\overline{f})^+|g^+\rangle-\langle (\overline{g})^+|f^+\rangle\right)(-1)^{N},
\end{equation}
vanishes for spacelike separated supports of the test functions $f$ and $g$. More precisely, $\langle (\overline{f})^+|g^+\rangle-\langle (\overline{g})^+|f^+\rangle$ equals zero for spacelike separation of the supports of $f$ and $g$.\par
We shall denote by $\widehat{\mathcal{R}}:W\mapsto\widehat{\mathcal{R}}(W)$ the net generated by the field $\widehat{\phi}$. Thus, in terms of the two $\mathcal{P}_+^{\uparrow}$-covariant nets $\{\mathcal{R}(W)\}_{W\in\mathcal{W}}$ and $\{\widehat{\mathcal{R}}(W)\}_{W\in\mathcal{W}}$ relative locality is expressed by
\begin{equation*}
\mathcal{R}(W)\subset\widehat{\mathcal{R}}(W')'=(-1)^{N(N-1)/2}\mathcal{R}(W')'(-1)^{N(N-1)/2}.
\end{equation*}

\subsection{Modular structure}
The analysis in \cite{BS1} revealed that the vacuum vector $\Omega$ is cyclic and separating for the algebras $\mathcal{R}(W)$ and $\widehat{\mathcal{R}}(W)$, where $W$ is any wedge region. Thus it is possible to determine the modular objects associated with the pairs $(\mathcal{R}(W),\Omega)$ and $(\widehat{\mathcal{R}}(W),\Omega)$, $W\in\mathcal{W}$. It turns out that the modular objects corresponding to $(\mathcal{R}(W),\Omega)$ and those corresponding to $(\widehat{\mathcal{R}}(W),\Omega)$ coincide. The modular operator and conjugation are given by
\begin{equation}\label{modulardata}
\Delta_W=U(\Lambda_W(2i\pi))\quad \mathrm{and}\quad J_W=U(j_W)
\end{equation}
respectively, where $\Lambda_W(t)$, $t\in\mathbb{R}$, is the one-parameter group of Lorentz boosts which leave the wedge $W$ invariant and $j_W$ is the reflection across the edge of the wedge $W$. The operator $U(j_W)$ acts according to
\begin{equation}\label{anti}
\left(U(j_W)\Psi\right)_n(p_1,\dots,p_n):=\overline{\Psi_n(-j_Wp_n,\dots,-j_Wp_1)}
\end{equation}
and extends the representation $U$ of $\mathcal{P}_+^{\uparrow}$ to a representation of $\mathcal{P}_+$. Moreover, we have $\mathcal{R}(W)'=\widehat{\mathcal{R}}(W')$.\par
In this setting the modular groups act geometrically correctly as expected from the Bisognano-Wichmann theorem, but as the model is not local the condition of geometric modular action \cite{BDFS} is not satisfied, i.e. the modular conjugations do not act geometrically correctly.

\subsection{The 2-dimensional case}\label{twodimundef}
The restriction to the two dimensional Minkowski spacetime allows for certain tools \cite[and papers quoted therein]{BL4} for analyzing the content of local observables of the model under consideration. Making use of these techniques, it is possible to show that the model at hand does contain nontrivial operators localized in double cones \cite{BL4, BS1, GL1}. Before giving any details, we start by noting that in $d=2$ wedge-locality can be implemented by defining
\begin{subequations}\label{net}
\begin{eqnarray}
\widetilde{\mathcal{R}}(W_0+x)&:=&\{\phi(f):f\in\mathscr{S}(W_0+x)\}'',\\
\widetilde{\mathcal{R}}(W_0'+x)&:=& \{\widehat{\phi}(f):f\in\mathscr{S}(W_0'+x)\}'',
\end{eqnarray}
\end{subequations}
where $x\in\mathbb{R}^2$. Due to the properties of the fields $\phi$ and $\widehat{\phi}$ it is clear from this definition that the resulting net $\{\widetilde{\mathcal{R}}(W)\}_{W\in\mathcal{W}}$ is wedge-local and transforms covariantly under Poincaré transformations. In more than two dimensions, however, this approach is not meaningful because one could rotate $W_0$ into $W_0'$ and obtain by covariance an algebra $\widetilde{\mathcal{R}}(W_0')$ generated by the field $\phi$. But as already discussed above $[\phi(f),\phi(g)]$ does not vanish at spacelike distances. The 2-dimensional case is special because there are no rotations mapping $W_0$ to $W_0'$.\par
In fact, within the 2-dimensional setting induced by Definition (\ref{net}) both the modular groups and the modular conjugation $J$ act geometrically correctly. Moreover, Haag duality holds, i.e. $\widetilde{\mathcal{R}}(W)'=\widetilde{\mathcal{R}}(W')$, $W\in\mathcal{W}$, \cite{BL4, GL2}.\par
As already mentioned above, it is known that the net (\ref{net}) contains nontrivial operators localized in bounded spacetime regions, namely double cones $\mathcal{O}$. More precisely, the local algebras $\mathcal{A}(\mathcal{O}):=\widetilde{\mathcal{R}}(W')\cap \widetilde{\mathcal{R}}(W+x)$, $\mathcal{O}:=W'\cap(W+x)$, $x\in W'$, have cyclic vectors and therefore contain nontrivial operators \cite{BL4, BS1, GL1}. In particular, the vacuum $\Omega$ is cyclic for the covariant and local net $\mathcal{A}$ and therefore the Haag-Ruelle-Hepp scattering theory is applicable. It turns out that the net $\mathcal{A}$ describes a Boson with nontrivial scattering matrix $S=(-1)^{N(N-1)/2}$. In particular, $S$ is factorizing and corresponds to the two-particle scattering function $S_2=-1$ \cite{GL2, GL3}.

\section{The deformed fermionic model}\label{3}
The deformation method presented in \cite{GL} yields a class of integrable models with factorizing S-matrices in two space-time dimensions \cite{AAR}. The S-matrix of such a model is completely determined by the two-particle scattering function $S_2$. The mentioned class of integrable models arises from deformation of a covariant local free quantum field theory and corresponds to scattering functions with value $+1$ at zero rapidity parameter, i.e. $S_2(0)=+1$. Models with scattering functions $S_2$ satisfying $S_2(0)=-1$, however, are not obtained in this way. This section is therefore devoted to the incorporation of these models into the deformation framework by deforming the model presented in Section \ref{model}. At the same time, our analysis complements the results in \cite{GL}.
\subsection{The deformation procedure}\label{deformiert}
We shall work within the framework introduced in Section \ref{model} and shall consider any spacetime dimension $d\geq 2$. Motivated by the deformation methods presented in \cite{GG} and \cite[Chap. 4]{GL}, our deformation approach involves first of all an operator-valued function $T_R:\mathbb{R}^{d}\rightarrow \mathcal{B}(\mathscr{H})$ which is defined by
\begin{equation}\label{TR}
\left(T_{R}(x)\Psi\right)_{n}(p_{1},\dots,p_{n}):=\prod_{k=1}^{n}R(x\cdot p_{k})\Psi_{n}(p_{1},\dots,p_{n}),
\end{equation}
with $\Psi\in\mathscr{H}$. The function $R$, hereinafter referred to as the deformation function, should satisfy the following conditions
\begin{definition}\label{R}
A deformation function is a continuous function $R:\mathbb{R}\rightarrow\mathbb{C}$ such that the following properties hold:
\begin{description}
    \item[i)]
\begin{equation*}
R(a)^{-1}=\overline{R(a)}
\end{equation*}
\item[ii)] The Fourier transform $\widetilde{R}$ of $R$ is a tempered distribution, i.e. $\widetilde{R}\in\mathscr{S}'$, and has support in $\mathbb{R}_+$, implying that $R$ extends to an analytic function on the upper half plane.
\item[iii)] The extension of $R$ to an analytic function on the upper half plane is continuous on the closure of the upper half plane.
\end{description}
\end{definition}
Note that the first property in Definition \ref{R} yields that $R(a)$ is a phase factor, i.e. $\left|R(a)\right|=1$. Therefore, $T_R(x)$ is a unitary operator, i.e. $T_R(x)^{*}=T_R(x)^{-1}$, since by Definition (\ref{TR}) we have
\begin{equation*}
T_R(x)^{*}=T_{\overline{R}}(x),\qquad T_R(x)^{-1}=T_{R^{-1}}(x).
\end{equation*}
The requirements ii) on the Fourier transform $\widetilde{R}$ of $R$ in Definition \ref{R} imply that $R$ extends to an analytic function on the upper half plane due to Theorem IX.16 in \cite{RS}. In particular, it follows from condition ii) that $R$ is the boundary value in the sense of $\mathscr{S}'$ of a function which is holomorphic in the upper half plane and satisfies polynomial bounds at infinity and at the real boundary. Condition iii) requires that the boundary value is even obtained in the sense of continuous functions.\par
Definition (\ref{TR}) further leads to the conclusion that for arbitrary deformation functions $R$ and $R'$ we have
\begin{equation}\label{multiplikativ}
T_{R}(x)T_{R'}(x)=T_{RR'}(x).
\end{equation}

In addition, we introduce a $(d\times d)$-matrix $Q$ which is antisymmetric w.r.t. the Minkowski inner product on $\mathbb{R}^{d}$ and satisfies
\begin{equation}\label{Q}
\Lambda Q\Lambda^{-1}=\left\{\begin{array}{ccc}
Q&\mathit{for}&\Lambda\in\mathcal{L}_+^{\uparrow}\,\,\mathit{with}\,\, \Lambda W_0=W_0\\
-Q&\mathit{for}&\Lambda\in\mathcal{L}_+^{\downarrow}\,\,\mathit{with}\,\, \Lambda W_0=W_0.
\end{array}\right.
\end{equation}
The most general $Q$ satisfying $(\ref{Q})$ is known to be of the form \cite{GG}
\begin{equation}\label{Qform}
Q = 
\left( \begin{array}{cccc}
0 & \kappa &0&0  \\
\kappa & 0  &0&0\\
0&0&0&\kappa '\\
0&0&-\kappa '&0
\end{array} \right),\qquad 
Q = 
\left( \begin{array}{ccccc}
0 & \kappa &0&\cdots&0  \\
\kappa & 0  &0&\cdots&0\\
\vdots&\vdots&\vdots &\ddots&\vdots\\
0&0&0&\cdots &0
\end{array} \right),
\end{equation}
for $d=4$ and $d\neq 4$ respectively and with $\kappa,\kappa '\in \mathbb{R}$. Moreover, we have
\begin{equation}\label{Q'}
\Lambda Q\Lambda^{-1}=\left\{\begin{array}{ccc}
-Q&\mathit{for}&\Lambda\in\mathcal{L}_+^{\uparrow}\,\,\mathit{with}\,\, \Lambda W_0=W_0'\\
Q&\mathit{for}&\Lambda\in\mathcal{L}_+^{\downarrow}\,\,\mathit{with}\,\, \Lambda W_0=W_0'.
\end{array}\right.
\end{equation}
Having introduced the necessary notation, we may now define deformed versions of the operator-valued distributions $a^{\#}(p)$ by
\begin{equation}\label{def1}
a^{*}_{R,Q}(p):=a^{*}(p)T_{R}(Qp)^{*}, \qquad a_{R,Q}(p):=a^{*}_{R,Q}(p)^{*}.
\end{equation}
We shall need the commutation relations of $a^{\#}(p)$ and $T_{R}(x)$, which can be computed very easily. First,
\begin{equation}\label{comm}
a(p)T_{R}(x)=R(x\cdot p)T_{R}(x)a(p),
\end{equation}
which for $x=Qp$ yields that $a(p)T_{R}(Qp)=R(0)T_{R}(Qp)a(p)$ due to the antisymmetry of the matrix $Q$. Taking adjoints, we find from equation (\ref{comm})
\begin{equation*}
a^{*}(p)T_{R}(x)^{*}=\overline{R(x\cdot p)}^{-1}T_{R}(x)^{*}a^{*}(p),
\end{equation*}
respectively
\begin{equation}\label{comm2}
a^{*}(p)T_{R}(x)=R(x\cdot p)^{-1}T_{R}(x)a^{*}(p).
\end{equation}
The deformed creation and annihilation operators therefore satisfy the following exchange relations for arbitrary $Q$ and $Q'$
\begin{subequations}\label{exchange}
\begin{eqnarray}
a^{*}_{R,Q}(p)a^{*}_{R,Q'}(q)&=&-\frac{R(Q'q\cdot p)}{R(Qp\cdot q)}a^{*}_{R,Q'}(q)a^{*}_{R,Q}(p),\\
a_{R,Q}(p)a_{R,Q'}(q)&=&-\frac{R(Q'q\cdot p)}{R(Qp\cdot q)}a_{R,Q'}(q)a_{R,Q}(p),
\end{eqnarray}
\begin{multline}
a_{R,Q}(p)a^{*}_{R,Q'}(q)\\
=\omega(\mathbf{p})\delta(\mathbf{p}-\mathbf{q}) T_{R}(Qp)T_{R}(Q'p)^{*}-
\frac{R(Qp\cdot q)}{R(Q'q\cdot p)}a^{*}_{R,Q'}(q)a_{R,Q}(p).
\end{multline}
\end{subequations}
Thus, as expected, the deformation has changed the underlying algebraic structure.\par
We may now introduce as usual corresponding field operators using the deformed creation and annihilation operators. These deformed field operators $\phi_{R,Q}(f)$ are defined by
\begin{equation}\label{def}
\phi_{R,Q}(f):=a^{*}_{R,Q}(f^{+})+a_{R,Q}(\overline{f^{-}}), \qquad f\in \mathscr{S}(\mathbb{R}^{d}),
\end{equation}
where for $\varphi\in\mathscr{H}_1$
\begin{equation*}
a_{R,Q}(\varphi)=\int d\mu(p) \overline{\varphi(p)}a_{R,Q}(p), \qquad a^{*}_{R,Q}(\varphi)=\int d\mu(p) \varphi(p)a^{*}_{R,Q}(p).
\end{equation*}
Note that if we set the deformation function $R(a)=-1$ for all $a\in\mathbb{R}$, the correspondingly deformed field operators are equal to the auxiliary fields given by Equation (\ref{hat}), i.e.
\begin{equation}\label{R-1}
\phi_{-1}(f)=\widehat{\phi}(f).
\end{equation}
For $R(a)=1$ for all $a\in\mathbb{R}$, one recovers the undeformed field $\phi$ given by (\ref{feld}), i.e. $\phi_{1}(f)=\phi(f)$.\par
In the same way as in the undeformed case, see Equation (\ref{hat}), we may also consider the auxiliary fields
\begin{equation}\label{phihut}
\widehat{\phi}_{R,Q}(f):=
(-1)^{N(N-1)/2}\phi_{R,Q}(f)(-1)^{N(N-1)/2}.
\end{equation}
Due to (\ref{multiplikativ}) and (\ref{R-1}), however, we have
\begin{equation}\label{phihut2}
\widehat{\phi}_{R,Q}(f)=\phi_{-R,Q}(f).
\end{equation}
In particular, in analogy to (\ref{R-1}) it follows
\begin{equation}\label{R-2}
\widehat{\phi}_{-1}(f)=\phi_{1}(f)=\phi(f).
\end{equation}
Due to the unitary equivalence
\begin{equation*}
a^{\#}_{-1}(\varphi)=(-1)^{N(N-1)/2}a^{\#}(\varphi)(-1)^{N(N-1)/2},\qquad \varphi\in\mathscr{H}_1,
\end{equation*}
the operator-valued distributions $a^{\#}_{-1}(p)$ also satisfy canonical anticommutation relations. Furthermore, it is straightforward to check that
\begin{subequations}\label{1}
\begin{equation}
[a(p), a_{-1}(q)] = 0,\qquad [a^{*}(p), a^{*}_{-1}(q)] = 0
\end{equation}
\begin{equation}
[a(p), a^{*}_{-1}(q)] = \{a(p), a^{*}(q)\}(-1)^{N}.
\end{equation}
\end{subequations}
\subsection{Properties of the deformed model in $d\geq 2$}\label{resultate}
In the following discussion we are interested in the features of the deformed field operators $\phi_{R,Q}(f)$. To begin with, we investigate domain and hermiticity properties, the Reeh-Schlieder property and the Klein-Gordon equation. Our results are given by the following proposition.
\begin{proposition}\label{prop1}
Let $R$ be a deformation function in the sense of Definition \ref{R} and let $Q$ be a $(d\times d)$-matrix which is antisymmetric w.r.t. the Minkowski inner product on $\mathbb{R}^{d}$ and satisfies (\ref{Q}) and (\ref{Q'}). Then the deformed field operators $\phi_{R,Q}(f)$, $f\in\mathscr{S}(\mathbb{R}^{d})$, have the following properties:
\begin{description}
    \item[a)] The dense subspace $\mathcal{D}\subset\mathscr{H}$ of vectors of finite particle number is contained in the domain $\mathcal{D}_{0}$ of any $\phi_{R,Q}(f)$. Moreover, $\phi_{R,Q}(f)\mathcal{D}\subset\mathcal{D}$ and $\phi_{R,Q}(f)\Omega=\phi(f)\Omega$.
    \item[b)] For $\Psi\in\mathcal{D}$ we have
\begin{equation}
\phi_{R,Q}(f)^{*}\Psi=\phi_{R,Q}(\overline{f})\Psi,
\end{equation}
and $\phi_{R,Q}(f)$ is essentially selfadjoint on $\mathcal{D}$ for real $f\in\mathscr{S}(\mathbb{R}^{d})$.
    \item[c)] $\phi_{R,Q}$ is a weak solution of the Klein-Gordon equation, i.e.
\begin{equation}
\phi_{R,Q}\left(\left(\Box+m^2\right)f\right)=0.
\end{equation}
    \item[d)] The Reeh-Schlieder property holds: For any non-empty open $\mathcal{O}\subset\mathbb{R}^d$ the set 
\begin{equation}
\mathcal{D}_{R,Q}(\mathcal{O}):=span\{\phi_{R,Q}(f_1)\cdots\phi_{R,Q}(f_n)\Omega:n\in\mathbb{N}_0, f_1,\dots,f_n\in\mathscr{S}(\mathcal{O})\}
\end{equation}
is dense in $\mathscr{H}$.
\end{description}
\end{proposition}
\textbf{\textit{Proof}}\\

a) These statements are a direct consequence of the definition of $\phi_{R,Q}$ (\ref{def}).\par
b) Since $\left(\overline{f}\right)^{\pm}=\overline{f^{\mp}}$ we have  $\phi_{R,Q}(f)^{*}\Psi=\phi_{R,Q}(\overline{f})\Psi$, $\Psi\in\mathcal{D}$. Along the same lines as \cite[Prop. 5.2.3]{BR} one can show the essential selfadjointness for real $f$. In particular, due to $R$ being a phase factor, we find for $\Psi_n\in\mathscr{H}_n$ the estimate
$$\|\phi_{R,Q}(f)\Psi_n\|\leq \left(\|f^+\|+\|f^-\|\right)\|(N+1)^{1/2}\Psi_n\|.$$
Therefore, for $k\in\mathbb{N}$
\begin{multline*}
\|\phi_{R,Q}(f)^k\Psi_n\|\leq(n+k)^{1/2}\left(\|f^+\|+\|f^-\|\right)\|\phi_{R,Q}(f)^{k-1}\Psi_n\|\leq\\
(n+k)^{1/2}\cdots(n+1)^{1/2}\left(\|f^+\|+\|f^-\|\right)^k\|\Psi_n\|.
\end{multline*}
This yields for arbitrary $t\in\mathbb{C}$ that
$$\sum_{k=0}^{\infty}|t|^k\frac{\|\phi_{R,Q}(f)^k\Psi_n\|}{k!}\leq\sum_{k=0}^{\infty}
\left(\frac{(n+k)!}{n!}\right)^{1/2}\frac{|t|^k}{k!}
\left(\|f^+\|+\|f^-\|\right)^k\|\Psi_n\|<\infty,$$
implying that every $\Psi\in\mathcal{D}$ is an analytic vector for $\phi_{R,Q}(f)$. Since $\mathcal{D}$ is dense in $\mathscr{H}$ and $\phi_{R,Q}(f)$ is hermitian for real $f$ one can apply Nelson's theorem \cite[Thm. X.39]{RS} and conclude that for real $f$, $\phi_{R,Q}(f)$ is essentially selfadjoint on $\mathcal{D}$.\par
c) This follows directly from $\left((\Box+m^2)f\right)^{\pm}=0$.\par
d) In order to prove this statement we want to make use of the spectrum condition and show in a standard manner \cite{SW} that $\mathcal{D}_{R,Q}(\mathcal{O})$ is dense in $\mathscr{H}$ if and only if $\mathcal{D}_{R,Q}(\mathbb{R}^d)\subset\mathscr{H}$ is dense. Thus, let $f_i\in\mathscr{S}(\mathbb{R}^d)$, $i=1,\dots,n$, with supp$\widetilde{f}_i\subset V_+$, then $\mathcal{D}_{R,Q}(\mathbb{R}^d)$ contains the vectors
$$\phi_{R,Q}(f_1)\cdots\phi_{R,Q}(f_n)\Omega=a_{R,Q}^*(f^+_1)\cdots a_{R,Q}^*(f^+_n)\Omega=\sqrt{n!}P_n\left(D_n(f_1^+\otimes\cdots\otimes f_n^+)\right),$$
where $P_n$ is the orthogonal projection from the unsymmetrized $\mathscr{H}_1^{\otimes n}$ onto its totally antisymmetric subspace $\mathscr{H}_n$, and $D_n\in\mathcal{B}(\mathscr{H}_1^{\otimes n})$ is the unitary operator multiplying with
$$D_n(p_1,\dots,p_n)=\prod_{1\leq k<l\leq n}R(Qp_k\cdot p_l)^{-1}.$$
By varying the test functions $f_i\in\mathscr{S}(\mathbb{R}^d)$ within this setting we obtain dense sets of $f^+_i$ in $\mathscr{H}_1$. Moreover, due to the unitary of $D_n$ this also leads to a total set of vectors $D_n(f_1^+\otimes\cdots\otimes f_n^+)$ in $\mathscr{H}_1^{\otimes n}$, implying that under the projection $P_n$ this set is total in $\mathscr{H}_n$. Hence it follows that $\mathcal{D}_{R,Q}(\mathbb{R}^d)$ is dense in $\mathscr{H}$. Application of the standard Reeh-Schlieder argument \cite{SW} finishes the proof.\hfill $\Box$\\

Furthermore, we are interested in the transformation behavior of the deformed fields $\phi_{R,Q}$ under the adjoint action of the representation $U$ of the Poincaré group $\mathcal{P}_+$. We find the following results.
\begin{lemma}\label{lem1}
The operator-valued function $T_R(Qp)$ defined by (\ref{TR}) transforms under the adjoint action of the representation $U$ of $\mathcal{P}_+$ (\ref{unitary}), (\ref{anti}) according to
\begin{subequations}
\begin{eqnarray}
U(a,\Lambda)T_{R}(Qp)U(a,\Lambda)^{-1}&=&T_{R}\left(\left(\Lambda Q\Lambda^{-1}\right)\Lambda p\right),\quad(a,\Lambda)\in \mathcal{P}_{+}^{\uparrow}\\
U(a,\Lambda)T_{R}(Qp)U(a,\Lambda)^{-1}&=&T_{R}\left(-\left(\Lambda Q\Lambda^{-1}\right)\Lambda p\right)^{*},\quad(a,\Lambda)\in \mathcal{P}_{+}^{\downarrow},
\end{eqnarray}
\end{subequations}
where $Q$ is a $(d\times d)$-matrix which is antisymmetric w.r.t. the Minkowski inner product on $\mathbb{R}^{d}$, satisfying (\ref{Q}) and (\ref{Q'}). Correspondingly, the operator-valued distributions $a^{\#}_{R,Q}(p)$ transform as follows
\begin{subequations}\label{UaU}
\begin{eqnarray}
U(a,\Lambda)a^{*}_{R,Q}(p)U(a,\Lambda)^{-1}&=&e^{i\Lambda p\cdot a}a^{*}_{R,\Lambda Q\Lambda^{-1}}(\Lambda p),\quad(a,\Lambda)\in \mathcal{P}_{+}^{\uparrow},\\
U(a,\Lambda)a^{*}_{R,Q}(p)U(a,\Lambda)^{-1}&=&e^{-i\Lambda p\cdot a}a^{*}_{-\overline{R},\Lambda Q\Lambda^{-1}}(-\Lambda p),\quad(a,\Lambda)\in \mathcal{P}_{+}^{\downarrow},
\end{eqnarray}
\end{subequations}
\begin{subequations}\label{UaU2}
\begin{eqnarray}
U(a,\Lambda)a_{R,Q}(p)U(a,\Lambda)^{-1}&=&e^{-i\Lambda p\cdot a}a_{R,\Lambda Q\Lambda^{-1}}(\Lambda p),\quad(a,\Lambda)\in \mathcal{P}_{+}^{\uparrow},\\
U(a,\Lambda)a_{R,Q}(p)U(a,\Lambda)^{-1}&=&e^{i\Lambda p\cdot a}a_{-\overline{R},\Lambda Q\Lambda^{-1}}(-\Lambda p),\quad(a,\Lambda)\in \mathcal{P}_{+}^{\downarrow}.
\end{eqnarray}
\end{subequations}
The smeared field operators $\phi_{R,Q}(f)$, $f\in\mathscr{S}(\mathbb{R}^{d})$, (\ref{def}) therefore satisfy
\begin{subequations}\label{UphiU}
\begin{eqnarray}
U(a,\Lambda)\phi_{R,Q}(f)U(a,\Lambda)^{-1}&=&\phi_{R,\Lambda Q\Lambda^{-1}}(f_{(a,\Lambda)}),\quad(a,\Lambda)\in \mathcal{P}_{+}^{\uparrow}\\
U(a,\Lambda)\phi_{R,Q}(f)U(a,\Lambda)^{-1}&=&\phi_{-\overline{R},\Lambda Q\Lambda^{-1}}(\overline{f}_{(a,\Lambda)}),\quad(a,\Lambda)\in \mathcal{P}_{+}^{\downarrow},
\end{eqnarray}
\end{subequations}
where $f_{(a,\Lambda)}(x)=f(\Lambda^{-1}(x-a))$.
\end{lemma}
\textbf{\textit{Proof}}\\\\
If $(a,\Lambda)\in\mathcal{P}_{+}^{\uparrow}$ and $\Psi\in\mathscr{H}$, then
\begin{eqnarray*}
\left(U(a,\Lambda)T_{R}(Qp)U(a,\Lambda)^{-1}\Psi\right)_n(p_1,\dots,p_n)&=&\prod_{k=1}^{n}R(Qp\cdot \Lambda^{-1} p_k)\Psi_n(p_1,\dots,p_n)\\
&=&\prod_{k=1}^{n}R(\Lambda Q\Lambda^{-1}\Lambda p\cdot  p_k)\Psi_n(p_1,\dots,p_n)\\
&=&\left(T_R(\Lambda Q\Lambda^{-1}\Lambda p)\Psi\right)_n(p_1,\dots,p_n),
\end{eqnarray*}
proving the first statement. Since $U(a,\Lambda)a(p)U(a,\Lambda)^{-1}=e^{-i\Lambda p\cdot a}a(\Lambda p)$ it follows for $a_{R,Q}(p)$
\begin{eqnarray*}
U(a,\Lambda)a_{R,Q}(p)U(a,\Lambda)^{-1}&=&e^{-i\Lambda p\cdot a}T_R(\Lambda Q\Lambda^{-1}\Lambda p)a(\Lambda p)\\
&=&e^{-i\Lambda p\cdot a}a_{R,\Lambda Q\Lambda^{-1}}(\Lambda p).
\end{eqnarray*}
Analogously, one shows the corresponding statement for $a_{R,Q}^{*}(p)$. For $(a,\Lambda)\in\mathcal{P}_{+}^{\downarrow}$ one finds
\begin{eqnarray*}
\left(U(a,\Lambda)T_{R}(Qp)U(a,\Lambda)^{-1}\Psi\right)_n(p_1,\dots,p_n)&=&\prod_{k=1}^{n}\overline{R(-Qp\cdot\Lambda^{-1} p_k)}\Psi_n(p_1,\dots,p_n)\\
&=&\left(T_R(-\Lambda Q\Lambda^{-1}\Lambda p)^{*}\Psi\right)_n(p_1,\dots,p_n).
\end{eqnarray*}
Hence with $U(a,\Lambda)a(p)U(a,\Lambda)^{-1}=e^{i\Lambda p\cdot a}a_{-1}(-\Lambda p)$ it follows
\begin{eqnarray*}
U(a,\Lambda)a_{R,Q}(p)U(a,\Lambda)^{-1}&=&e^{i\Lambda p\cdot a}T_{\overline{R}}(-\Lambda Q\Lambda^{-1}\Lambda p)a_{-1}(-\Lambda p)\\
&=&e^{i\Lambda p\cdot a}a_{-\overline{R},\Lambda Q\Lambda^{-1}}(-\Lambda p).
\end{eqnarray*}
For $a_{R,Q}^{*}(p)$ one proceeds in the same way. The transformation behavior (\ref{UphiU}) of the field $\phi_{R,Q}$ is a direct consequence of Equations (\ref{UaU}) and (\ref{UaU2}).\hfill $\Box$\\

The previous lemma shows that in the deformed model $\mathcal{P}_+$-covariance is violated. The property of $\mathcal{P}_+^{\uparrow}$-covariance is, however, preserved. To this end, let $\mathscr{P}_{R}(W_0)$ denote the polynomial algebra of fields generated by all $\phi_{R,Q}(f)$ with $f\in\mathscr{S}(W_0)$. It follows from the transformation behavior (\ref{UphiU}) that the algebra
\begin{equation}
\mathscr{P}_{R}(\Lambda W_0+a):=U(a,\Lambda)\mathscr{P}_{R}(W_0)U(a,\Lambda)^{-1}, \qquad (a,\Lambda)\in\mathcal{P}_+^{\uparrow}
\end{equation}
is generated by the fields $\phi_{R,\Lambda Q\Lambda^{-1}}(f)$ with $f\in\mathscr{S}(\Lambda W_0+a)$, and the corresponding net $W\mapsto \mathscr{P}_{R}(W)$, $W\in\mathcal W$, is $\mathcal{P}_+^{\uparrow}$-covariant.\par
In two spacetime dimensions the deformed theory admits a $\mathcal{P}_+$-covariant net if an additional condition is imposed on the deformation function $R$, see Section \ref{2}.\par
Note that there is a connection between the set of wedges $\mathcal{W}$ and the orbit $\mathcal{Q}:=\{\Lambda Q\Lambda^{-1}:\Lambda\in\mathcal{L}_+\}$. Namely, $\mathcal{Q}$ is in one-to-one correspondence with wedges whose edges contain the origin \cite{GG}. The deformation function $R$, on the other hand, specifies the kind of deformation that is used.\\

It is clear that in general the properties of the deformed field $\phi_{R,Q}$ differ from those of the undeformed field $\phi$. In particular, $\phi$ is a bounded operator, whereas $\phi_{R,Q}$ is in general not as the exchange relations (\ref{exchange}) imply. $\phi$ and $\phi_{R,Q}$, however, have in common that they are both nonlocal fields. The nonlocality of $\phi_{R,Q}$ can be explicitly seen by computing the two-particle contribution of the field commutator $[\phi_{R,Q}(f),\phi_{R,Q}(g)]$ applied to the vacuum $\Omega$, which yields
\begin{equation}\label{nloc}
\int d\mu(p)d\mu(q)f^{+}(p)g^{+}(q)\left(\overline{R(q\cdot Qp)}a^{*}(p)a^{*}(q)+\overline{R(p\cdot Qq)}a^{*}(p)a^{*}(q)\right)\Omega.
\end{equation}
This expression, however, only vanishes if $R(a)=-R(-a)$, $\forall a\in\mathbb{R}$. This requirement may be true for a function that fulfills $R(0)=0$, but with regard to Definition \ref{R} that requires $|R(a)|=1$ such a deformation function is inadmissible.\par
Note that in contrast to the deformation of a bosonic model \cite{GG,GL}, where $\phi_{R,Q}^{\rm{CCR}}(f)$ is relatively local to $\phi_{R,-Q}^{\rm{CCR}}(g)$, i.e. $[\phi_{R,Q}^{\rm{CCR}}(f),\phi_{R,-Q}^{\rm{CCR}}(g)]=0$ for supp$\,f\subset W_0$ and supp$\,g\subset W_0'$, $\phi_{R,Q}(f)$ is not relatively local to $\phi_{R,-Q}(g)$ for supp$\,f\subset W_0$ and supp$\,g\subset W_0'$. In particular, the two-particle contribution of $[\phi_{R,Q}(f),\phi_{R,-Q}(g)]$ applied to the vacuum reads
\begin{equation*}
2\int d\mu(p)d\mu(q)f^{+}(p)g^{+}(q)\overline{R(q\cdot Qp)}a^{*}(p)a^{*}(q)\Omega
\end{equation*}
which because of Definition \ref{R} does not vanish.\par
Along the lines of the undeformed case, see Equation (\ref{relloc}), we may consider the field commutator $[\phi_{R,Q}(f),\widehat{\phi}_{R,-Q}(g)]=[\phi_{R,Q}(f),\phi_{-R,-Q}(g)]$ for supp$\,f\subset W_0$ and supp$\,g\subset W_0'$. For the investigation of this commutator it is necessary to compute the corresponding commutation relations of the operators $a_{R,Q}^{\#}$ with $a_{-R,-Q}^{\#}$. A simple calculation shows that
\begin{subequations}\label{12}
\begin{equation}
[a_{R,Q}(p),a_{-R,-Q}(q)]=0,\qquad [a^{*}_{R,Q}(p),a^{*}_{-R,-Q}(q)]=0,
\end{equation}
\begin{equation}
[a_{R,Q}(p),a_{-R,-Q}^{*}(q)]
=\omega(\mathbf{p})\delta(\mathbf{p}-\mathbf{q})
(-1)^N T_{R}(Qp)T_{R}(-Qp)^{*},
\end{equation}
\begin{equation}
[a_{R,Q}^{*}(p),a_{-R,-Q}(q)] 
=\omega(\mathbf{p})\delta(\mathbf{p}-\mathbf{q})
(-1)^{N+1}T_{R}(Qp)^{*}T_{R}(-Qp).
\end{equation}
\end{subequations}
\begin{proposition}\label{co}
Let $R$ be a deformation function in the sense of Definition \ref{R} and $Q$ a $(d\times d)$-matrix which is antisymmetric w.r.t. the Minkowski inner product on $\mathbb{R}^{d}$, satisfying (\ref{Q}) and (\ref{Q'}). If $\kappa\geq 0$ in (\ref{Qform}), then the field operators $\phi_{R,Q}(f)$ (\ref{def}) and $\phi_{-R,-Q}(g)$ are relatively wedge-local to each other, i.e. for $f\in\mathscr{S}(W_0)$, $g\in\mathscr{S}(W_0')$
\begin{equation}\label{commuu}
[\phi_{R,Q}(f),\phi_{-R,-Q}(g)]\Psi=0,\qquad\Psi\in\mathcal{D},
\end{equation}
holds.
\end{proposition}
\textbf{\textit{Proof}}\\

Since $\langle\Phi,[\phi_{R,Q}(f),\phi_{-R,-Q}(g)]\Psi\rangle$, $\Phi, \Psi\in\mathcal{D}$, is a tempered distribution in $f$ and $g$, vanishing on $C_{0}^{\infty}(W_0)\times C_{0}^{\infty}(W_0')$ implies vanishing on $\mathscr{S}(W_0)\times\mathscr{S}(W_0')$. Making use of that property it thus suffices to prove (\ref{commuu}) for $(f,g)\in C_{0}^{\infty}(W_0)\times C_{0}^{\infty}(W_0')$.\par
Due to the commutation relations (\ref{12}) we have
\begin{multline*}
[\phi_{R,Q}(f),\phi_{-R,-Q}(g)]\Psi=\left([a_{R,Q}(\overline{f^-}),a_{-R,-Q}^*(g^+)]+
[a_{R,Q}^*(f^+),a_{-R,-Q}(\overline{g^-})]\right)\Psi,
\end{multline*}
which together with Definition \ref{R} yields the following n-particle contribution of this vector
\begin{multline}\label{int}
\left([\phi_{R,Q}(f),\phi_{-R,-Q}(g)]\Psi\right)_{n}(p_{1},\dots,p_{n})
\\=(-1)^{n}\int d\mu(p)\left(f^{-}(p)g^{+}(p)\prod_{k=1}^{n}\frac{R(p_{k}\cdot Qp)}{R(-p_{k}\cdot Qp)}-
f^{+}(p)g^{-}(p)\prod_{k=1}^{n}\frac{R(-p_{k}\cdot Qp)}{R(p_{k}\cdot Qp)}\right)\\
\times\Psi_{n}(p_{1},\dots,p_{n}).
\end{multline}
Our task is now to show that this expression vanishes for all $p_k$, $k=1,\dots,n$. Following the proof of Proposition 3.4 in \cite{GG} we may introduce new coordinates:
\begin{equation*}
m_{\perp}:=\sqrt{m^{2}+p_{\perp}^{2}},\qquad p_{\perp}:=(p_{2},\dots,p_{d-1}),\qquad \vartheta:=\mathrm{Arsinh} \frac{p_{1}}{m_{\perp}}.
\end{equation*}
Thus, in the coordinates $(\vartheta,p_{\perp})$ we have
\begin{equation*}
d\mu(p)=\frac{d^{d-1}\mathbf{p}}{\omega(\mathbf{p})}=d\vartheta d^{d-2}p_{\perp},
 \qquad p=p(\vartheta):=\left( \begin{array}{c}
m_{\perp}\rm{cosh}\,\vartheta   \\
m_{\perp}\rm{sinh}\,\vartheta \\
p_{\perp}
\end{array} \right).
\end{equation*}
Correspondingly, we use the following notation
$$f^{\pm}(\vartheta,p_{\perp}):=\widetilde{f}(\pm p(\vartheta)).$$
According to \cite{GG}, $f^{-}(\vartheta+i\lambda,p_{\perp})$ is bounded on the strip $0\leq\lambda\leq\pi$, $\vartheta\in\mathbb{R}$, due to supp$f\subset W_0$ and analyticity properties of $\widetilde{f}$, $f\in C_{0}^{\infty}(W_0)$. In particular, $\widetilde{f}$ is an entire analytic function because $f$ has compact support. Moreover, also $g^{+}(\vartheta+i\lambda,p_{\perp})$, supp$\,g\subset W_0'=-W_0$, is bounded on the strip $0\leq\lambda\leq\pi$, $\vartheta\in\mathbb{R}$, and the boundary values at $\lambda=\pi$ are given by
\begin{equation}\label{perp}
f^{-}(\vartheta+i\pi,p_{\perp})=f^{+}(\vartheta,-p_{\perp}),\qquad
g^{+}(\vartheta+i\pi,p_{\perp})=g^{-}(\vartheta,-p_{\perp}).
\end{equation}
It remains to study the properties of the functions $\vartheta\mapsto R(Qp(\vartheta)\cdot p_k)\overline{R(-Qp(\vartheta)\cdot p_k)}$, $k=1,\dots,n$, which appear in (\ref{int}). It follows for $0\leq\lambda\leq\pi$ that
$$\mathrm{Im}\left(p(\vartheta+i\lambda)Q\cdot p_k\right)=\kappa m_\perp \mathrm{sin}\,\lambda\left(\begin{array}{c}
\rm{cosh}\,\vartheta   \\
\rm{sinh}\,\vartheta 
\end{array} \right)\cdot \left(\begin{array}{c}
p_k^0   \\
p_k^1
\end{array} \right)\geq 0$$
because $\kappa\geq 0$ and both $(\rm{cosh}\,\vartheta,\rm{sinh}\,\vartheta)$ and $(p_k^0,p_k^1)$ are in the two-dimensional forward lightcone.
Due to Definition \ref{R}, this implies that the functions $z\mapsto R(Qp(z)\cdot p_k)\overline{R(-Qp(z)\cdot p_k)}$, $k=1,\dots,n$, are analytic on the strip $S(0,\pi):=\{z=\vartheta+i\lambda\in\mathbb{C}:0<\lambda<\pi\}$. In addition, it also follows from Definition \ref{R} that these functions are continuous on the closure $\overline{S(0,\pi)}$ of $S(0,\pi)$, which implies that $|R(Qp(z)\cdot p_k)\overline{R(-Qp(z)\cdot p_k)}|\leq 1$ for $z\in\overline{S(0,\pi)}$ \cite[Thm. 12.9]{Ru}.
Hence, together with the previous discussion it is possible to shift the $\vartheta$-integration in (\ref{int}) from $\mathbb{R}$ to $\mathbb{R}+i\pi$. Making use of (\ref{perp}), we have
\begin{eqnarray*}\label{rel}
&&\int d\mu(p)f^{-}(p)g^{+}(p)\prod_{k=1}^{n}\frac{R(p_{k}\cdot Qp)}{R(-p_{k}\cdot Qp)}\\
&=&\int d^{d-2}p_{\perp}\int d\vartheta f^{-}(\vartheta,p_{\perp})g^{+}(\vartheta,p_{\perp})
\prod_{k=1}^{n}\frac{R(p_{k}\cdot Qp(\vartheta))}{R(-p_{k}\cdot Qp(\vartheta))}\\
&=&\int d^{d-2}p_{\perp}\int d\vartheta f^{+}(\vartheta,-p_{\perp})g^{-}(\vartheta,-p_{\perp})\prod_{k=1}^{n}\frac{R(p_{k}\cdot Qp(\vartheta+i\pi))}{R(-p_{k}\cdot Qp(\vartheta+i\pi))}\\
&=&\int d\mu(p)f^{+}(p)g^{-}(p)\prod_{k=1}^{n}\frac{R(-p_{k}\cdot Qp)}{R(p_{k}\cdot Qp)}.
\end{eqnarray*}
Thus,
\begin{equation*}
\left([\phi_{R,Q}(f),\phi_{-R,-Q}(g)]\Psi\right)_{n}(p_{1},\dots,p_{n})=0
\end{equation*}
for supp$\,f\subset W_0$ and supp$\,g\subset W_0'$.\hfill $\Box$\\
\begin{corollary}
Let $R$ be a deformation function in the sense of Definition \ref{R} and $Q$ a $(d\times d)$-matrix which is antisymmetric w.r.t. the Minkowski inner product on $\mathbb{R}^{d}$, satisfying (\ref{Q}) and (\ref{Q'}). If $\kappa<0$ in (\ref{Qform}), then the field operators $\phi_{R,-Q}(f)$ (\ref{def}) and $\phi_{-R,Q}(g)$ are relatively wedge-local to each other, i.e. for $f\in\mathscr{S}(W_0)$, $g\in\mathscr{S}(W_0')$
\begin{equation}\label{commu}
[\phi_{R,-Q}(f),\phi_{-R,Q}(g)]\Psi=0,\qquad\Psi\in\mathcal{D},
\end{equation}
holds.
\end{corollary}
The proof of this statement is analogous to the one of Proposition \ref{co}.\\

As a consequence of Proposition \ref{co} the $\mathcal{P}_+^\uparrow$-covariant nets $\mathscr{P}_{R}$ and $\mathscr{P}_{-R}$ are relatively wedge-local in the sense that
\begin{equation*}
\mathscr{P}_{R}(W)\subset\mathscr{P}_{-R}(W')',
\end{equation*}
for $W=\Lambda W_0 +a$.
Hence, in analogy to the undeformed case, see Section \ref{model}, we are dealing with nonlocal nets. Due to Proposition \ref{prop1} it is, however, possible to proceed from the net $W\mapsto\mathscr{P}_{R}(W)$ to the corresponding net of von Neumann algebras $\mathscr{N}_{R}(W)$. Considering these nets of bounded operators, one may analyze intersections of algebras such as
\begin{equation*}
\mathcal{I}_{R}(W_1\cap W_2'):=\mathscr{N}_{R}(W_1)\cap \mathscr{N}_{-R}(W_2'),\qquad
\overline{W_2}\subset W_1,\quad W_1,W_2\in\mathcal{W},
\end{equation*}
For the special case $R=1$ the authors in \cite{BS1} show that such intersections are not trivial and the corresponding net fulfills certain locality and covariance properties, depending on the spacetime dimension $d$. For general deformation functions $R$ this is still an open problem which is being pursued.\\

Note that in contrast to the results in \cite{GL} where the deformation of a bosonic model is investigated, we did not require that the deformation function $R$ satisfies $R(a)^{-1}=R(-a)$ and $R(0)=1$. In particular, $R(0)=1$ in \cite{GL} results from the deformation of the underlying Borchers-Uhlmann algebra $\underline{\mathscr{S}}$. More precisely, the deformation is based on linear homeomorphisms $\rho:\underline{\mathscr{S}}\rightarrow\underline{\mathscr{S}}$ with $\rho(1)=1$ and $\rho(f)^*=\rho(f^*)$, $f\in\underline{\mathscr{S}}$, which endow $\underline{\mathscr{S}}$ with a new product $\otimes_\rho$ defined by
$$f\otimes_{\rho}g:=\rho^{-1}(\rho(f)\otimes\rho(g)),\qquad f,g\in\underline{\mathscr{S}}.$$
Requiring a certain compatibility between $\rho$ and a state $\omega$ on $\underline{\mathscr{S}}$, namely
\begin{equation}\label{komp}
\omega(f\otimes_{\rho}g)=\omega(f\otimes g),\qquad f,g\in\underline{\mathscr{S}},
\end{equation}
the representation spaces arising from GNS construction are identical for the deformed and undeformed case, simplifying the analysis. Moreover, assuming that the deformation maps $\rho$ act multiplicatively in momentum space, i.e.
$$\widetilde{\rho(f)}_n(p_1,\dots,p_n):=\rho_n(p_1,\dots,p_n)\cdot \widetilde{f}_n(p_1,\dots,p_n),$$
the compatibility requirement (\ref{komp}) for quasi-free, translationally invariant states $\omega$ yields explicit conditions on the functions $\rho_n\in C^\infty(\mathbb{R}^{nd})$, $n\in\mathbb{N}_0$. In particular, it turns out that the functions $\rho_n$ are determined by the functions $\rho_2$. The connection to our deformation approach is given by $$\rho_2(p,q):=R(-p\cdot Qq).$$
The conditions on $\rho_2$ yield, inter alia, $R(a)^{-1}=R(-a)$ and $R(0)=1$.\par 
However, in the deformed bosonic case \cite{GL} the requirement $R(a)^{-1}=R(-a)$ is necessary for obtaining wedge-locality and covariance. In contrast to this, we do not obtain the same result for our deformed fermionic model by imposing this relation, except in two spacetime dimensions, see Section \ref{2}. Nevertheless, the requirement $R(0)=1$ is redundant for establishing wedge-locality and covariance properties for the deformed model in both the deformed bosonic and the deformed fermionic case. In particular, one can perform a deformation as presented in Section \ref{deformiert} of a bosonic model involving field operators $\phi^{CCR}$ and arrive at a covariant and wedge-local deformed model involving deformed field operators $\phi_{R,Q}^{CCR}$ with deformation functions $R$ not necessarily satisfying $R(0)=1$ as is the case in \cite{GL}.\par
In fact, considering a deformation function $R$, the correspondingly deformed net $\mathscr{P}_{R}$ is unitarily equivalent to the net $\mathscr{P}_{-R}$, implying that deformations involving $R$ and those involving $-R$ are equivalent. In other words, a model resulting from deformation associated with $R$ is physically indistinguishable from a model arising from deformation with $-R$. We summarize this result in the following lemma.
\begin{lemma}\label{lem2}
The net $\mathscr{P}_{R}$ is unitarily equivalent to the net $\mathscr{P}_{-R}$. The unitary $V$ relating these two nets is given by
\begin{equation*}
V:=(-1)^{N(N-1)/2},
\end{equation*}
where $N$ is the particle number operator, i.e. $N|_{\mathscr{H}_n}=n\cdot 1$.
\end{lemma}
\textbf{\textit{Proof}}\\

Since the unitary $V$ commutes with all Poincaré transformations, i.e. 
$$[V,U(g)]=0\qquad\forall\, g\in\mathcal{P}_+,$$
and satisfies $V\Omega=\Omega$, the unitary equivalence $V\mathscr{P}_{R}V^{-1}=\mathscr{P}_{-R}$ follows.\hfill $\Box$

\subsection{The 2-dimensional case and integrable models}\label{2}
In this section we shall consider the case of a two-dimensional Minkowski space. Recall that in $d=2$ the set of wedges $\mathcal{W}$ consists of two disjoint subsets, namely the translates of $W_0$ (\ref{W}) and the translates of $W_0'=-W_0$. Moreover, the matrix $Q$ is of the form
\begin{equation}\label{Qfor}
Q = \lambda
\left( \begin{array}{cc}
0 & 1  \\
1 & 0  
\end{array} \right),\qquad \lambda\in \mathbb{R}.
\end{equation}
In analogy to the undeformed case (\ref{net}) we may define for a fixed deformation function $R$ given by Definition \ref{R} and an admissible and fixed matrix $Q$ (\ref{Qfor}) the polynomial algebras $\widetilde{\mathscr{P}}(W_0+x)$ and $\widetilde{\mathscr{P}}(W_0'+x)$, $x\in\mathbb{R}^2$, which are generated by the fields $\phi_{R,Q}(f)$, $f\in\mathscr{S}(W_0+x)$, and $\phi_{-\overline{R},Q}(f)$, $f\in\mathscr{S}(W_0'+x)$, respectively, i.e.
\begin{subequations}\label{netdeformed}
\begin{eqnarray}
\widetilde{\mathscr{P}}(W_0+x):=\{\mathrm{polynomials}\, \mathrm{in}\,\phi_{R,Q}(f):f\in\mathscr{S}(W_0+x)\},\\
\widetilde{\mathscr{P}}(W_0'+x):=\{\mathrm{polynomials}\, \mathrm{in}\,\phi_{-\overline{R},Q}(f):f\in\mathscr{S}(W_0'+x)\}.
\end{eqnarray}
\end{subequations}
This definition, however, only produces a wedge-local and covariant net $W\mapsto\widetilde{\mathscr{P}}(W)$, $W\in\mathcal{W}$, if the deformation function $R$ fulfills the property
\begin{equation}
\overline{R(a)}=R(-a),\qquad\forall\, a\in\mathbb{R}.
\end{equation}
If, namely, this property holds, the fields $\phi_{-\overline{R},Q}(f)$ are equal to $\phi_{-R,-Q}(f)$ and wedge-locality follows from Proposition \ref{co}. We shall therefore assume this relation in what follows. Due to Lemma \ref{lem1} the net $W\mapsto\widetilde{\mathscr{P}}(W)$, $W\in\mathcal{W}$, also transforms covariantly under the adjoint action of the representation $U$ of $\mathcal{P}_+$.\par
In the following, we would like to relate the covariant and wedge-local net $\{\widetilde{\mathscr{P}}(W)\}_{W\in\mathcal{W}}$ to an integrable quantum field theory model with factorizing S-matrix on two-dimensional Minkowski space. To this end, we start by noting that in two dimensions one may parametrize $H_{m}^{+}$ with the help of the rapidity $\theta\in\mathbb{R}$, i.e. $p(\theta):=m(\rm{cosh}\,\theta,\rm{sinh}\,\theta)$. Making use of this notation and (\ref{Qfor}), we have
\begin{equation*}
-p(\theta_{1})\cdot Qp(\theta_{2})=\lambda m^{2}\rm{sinh}(\theta_{1}-\theta_{2}),\qquad \theta_1,\theta_2\in\mathbb{R}.
\end{equation*}
We further define
\begin{equation}\label{scatteringfunction}
S_{\lambda}:\mathbb{R}\rightarrow\mathbb{C},\qquad S_{\lambda}(\theta):=-R(\lambda m^{2}\rm{sinh}\,\theta)^2.
\end{equation}
Since the entire analytic function sinh maps the strip $S(0,\pi):=\{z\in\mathbb{C}:0<\mathrm{Im}\,z<\pi\}$ onto the upper half plane and since by Definition \ref{R} $R$ has an analytic continuation to the upper half plane, the function $S_\lambda$, $\lambda\geq 0$, extends to an analytic function on the strip $S(0,\pi)$. Moreover, it follows from the requirements on the function $R$ by Definition \ref{R} and the properties of sinh that
\begin{equation*}
S_{\lambda}(0)=-1,\qquad \overline{S_{\lambda}(\theta)}=S_{\lambda}(-\theta)=S_{\lambda}(\theta)^{-1}=
S_{\lambda}(\theta+i\pi),\qquad \lambda,\theta\in\mathbb{R}.
\end{equation*}
These properties of the function $S_\lambda$ are familiar from the context of factorizing S-matrices and express the unitarity, hermitian analyticity and crossing symmetry of the scattering operator $S$ associated with $S_\lambda$ \cite{I, S}. In addition, the Yang-Baxter equation is trivially fulfilled by $S$ because we are considering here only a single species of particles. Due to these properties the scattering operator $S$ associated with $S_\lambda$ agrees with an S-matrix of a completely integrable relativistic quantum field theory \cite{S}.\par
The connection of an integrable quantum field theory model to the deformation procedure carried out in Section \ref{deformiert} and therefore to the net $\{\widetilde{\mathscr{P}}(W)\}_{W\in\mathcal{W}}$ may be clarified by introducing
\begin{equation*}
z_{\lambda}(\theta):=a_{R,Q}(p(\theta)),\qquad z^{\dagger}_{\lambda}(\theta):=a^{*}_{R,Q}(p(\theta)).
\end{equation*}
The exchange relations (\ref{exchange}) for $Q=Q'$ and $Q$ given by (\ref{Qfor}) then read
\begin{eqnarray*}
z_{\lambda}(\theta_{1})z_{\lambda}(\theta_{2})&=&S_{\lambda}(\theta_{2}-\theta_{1})
z_{\lambda}(\theta_{2})z_{\lambda}(\theta_{1})\\
z^{\dagger}_{\lambda}(\theta_{1})z^{\dagger}_{\lambda}(\theta_{2})&=&S_{\lambda}(\theta_{2}-\theta_{1})
z^{\dagger}_{\lambda}(\theta_{2})z^{\dagger}_{\lambda}(\theta_{1})\\
z_{\lambda}(\theta_{1})z^{\dagger}_{\lambda}(\theta_{2})&=&S_{\lambda}(\theta_{1}-\theta_{2}) z^{\dagger}_{\lambda}(\theta_{2})z_{\lambda}(\theta_{1})+\delta(\theta_{1}-\theta_{2})\cdot 1.
\end{eqnarray*}
That is, $z_{\lambda}(\theta)$ and $z^{\dagger}_{\lambda}(\theta)$ form a representation of the Zamolodchikov-Faddeev algebra \cite{F, Z} with scattering function $S_{\lambda}(\theta)$. Initiated by B. Schroer \cite{Sch}, it was shown \cite{BL4, GL03, GL3, Sch} that one can use this algebraic structure as a starting point for the construction of quantum field theories with factorizing S-matrices. Within this approach, one uses the fields $\phi_{\lambda}(x):=\int d\theta(e^{ip(\theta)\cdot x}z_\lambda^{\dagger}(\theta)+e^{-ip(\theta)\cdot x}z_\lambda(\theta))$ associated with $z^{\#}_{\lambda}(\theta)$ as wedge-local polarization-free generators for constructing model theories. The interesting point here is that  these fields appear in the present setting as a consequence of the deformation of the model given in Section \ref{model}. More precisely, the fields $\phi_{\lambda}$ coincide with the deformed fields $\phi_{R,Q}\in\widetilde{\mathscr{P}}$.\par
As already mentioned in Section \ref{twodimundef}, in $d=2$ there are certain operator-algebraic techniques by means of which it is possible to analyze the content of local observables of the considered model \cite{BL4}. Using these tools, it was shown in \cite{GL3} that if $S_\lambda$ is a regular scattering function in the sense that $z\mapsto S_\lambda(z)$ can be extended to a bounded analytic function on the strip $\{z\in\mathbb{C}:-\varepsilon<\mathrm{Im}\,z<\pi+\varepsilon\}$ for some $\varepsilon>0$, then the quantum field theory arising from $\phi_{\lambda}$ contains nontrivial observables localized in arbitrarily small open regions $\mathcal{O}\subset\mathbb{R}^2$. Moreover, besides other standard properties of quantum field theory also the Reeh-Schlieder property holds. In addition, the S-matrix of the model is found to be the one determined by the two-particle scattering function $S_\lambda$ \cite{GL3}. The following theorem demonstrates the connection of the deformation of a scalar massive Fermion to integrable models.
\begin{theorem}\label{theorem}
Every integrable quantum field theory on two-dimensional Minkowski space with scattering function $S_\lambda$ of the form (\ref{scatteringfunction}) can be obtained by deformation of a scalar massive Fermion in the sense of Section \ref{deformiert} provided the deformation function $R$ satisfies $R(-a)=R(a)^{-1}$ for all $a\in\mathbb{R}$. If further $S_\lambda$ is regular, then in the deformed theory there exist observables localized in double cones, and the Reeh-Schlieder property holds \cite[Thm. 5.8]{GL3}.
\end{theorem}
Thus, the analysis presented in this paper provides a complement to the results in $\cite{GL}$ as there the class of integrable models with scattering functions satisfying $S_\lambda(0)=-1$ was not obtained by means of deformation techniques.\par
Note that since the deformation function $R$ appears quadratically in the definition of the scattering function $S_\lambda$ (\ref{scatteringfunction}), $S_\lambda$ does not depend on the sign of $R$. This circumstance implies physical indistinguishability of correspondingly deformed models, i.e. models arising from deformation with $R$ and $-R$, which is in agreement with Lemma \ref{lem2}.\par
We close this section by giving concrete examples of deformation functions $R$ for which Theorem \ref{theorem} applies, namely
\begin{equation*}
R(a)=\pm\prod_{k=1}^n\frac{z_k-a}{z_k+a},\qquad \mathrm{Im}\,z_k> 0,
\end{equation*}
where for each $z_k$ also $-\overline{z_k}$ is contained in the set of zeros $\{z_1,\dots,z_n\}$.

\section{Conclusions and open questions}\label{conclusions}
Starting from a model of a scalar massive Fermion we included a certain class of integrable quantum field theory models into the deformation framework in two-dimensional Minkowski space. Namely, these are those integrable models whose factorizing S-matrices are completely determined by scattering functions $S_2$ satisfying $S_2(0)=-1$. For example, the scattering function of the Sinh-Gordon model belongs to this class. The analysis presented in this paper therefore provides a complement to the results in \cite{GL} where $S_2$ was required to satisfy $S_2(0)=1$.\par
The establishment of locality properties of the deformed model turns out to be a difficult task as the undeformed model is already nonlocal. In two dimensions, however, it is possible to achieve wedge-locality by imposing an additional condition on the deformation function $R$, namely $R(a)^{-1}=R(-a)$ for all $a\in\mathbb{R}$. Moreover, it follows from the analysis of integrable models  \cite{GL3} that the deformed theory also admits local observables in $d=2$. Analogous results for higher dimensions have not been achieved up to now. This problem is, however, under investigation.\par
One can also ask if some of the conditions on the deformation function $R$ can be relaxed. As part of our analysis, it turned out that from the physical point of view the deformed theory does not depend on the sign of the function $R$. In particular, two nets arising from deformation with deformation functions $R$ and $-R$ respectively are unitarily equivalent. This result generalizes the deformation procedure of \cite{GL} because there one requires that the function $R$ satisfies the condition $R(0)=1$ which by our result is redundant. Since the latter condition is a consequence of the deformation of the underlying Borchers-Uhlmann algebra \cite{GL}, the deformation approach presented here extends the possibilities for obtaining new models by deformation techniques.\par
It is expected from the simple form of the deformation that if our deformed model admits interaction in more than two dimensions, the theory will not involve momentum transfer or particle production and therefore will not be physically realistic. In order to realize these interactions, the generalization of deformation techniques is currently being developed and will be presented elsewhere.

\section*{Acknowledgments}
I would like to thank the Vienna deformation group and A. Much for helpful discussions and in particular, G. Lechner, J. Schlemmer and J. Yngvason who provided constant support.\par
This work was supported by FWF-project P22929-N16 "Deformations of Quantum Field Theories".

\end{document}